\documentclass{ws-procs9x6}

\begin{document}

\title{Present understanding of the \\
nucleon spin 
structure\footnote{\uppercase{T}his work has been supported by the 
\uppercase{S}ofia \uppercase{K}ovalevskaya \uppercase{P}rogramme 
of the \uppercase{A}lexander von \uppercase{H}umboldt \uppercase{F}oundation
and by the \uppercase{D}eutsche \uppercase{F}orschungsgemeinschaft.}}

\author{A. METZ}

\address{Institut f\"ur Theoretische Physik II, \\
Ruhr-Universit\"at Bochum, 44780 Bochum, Germany\\ 
E-mail: metza@tp2.ruhr-uni-bochum.de}

\maketitle

\abstracts{The present understanding of the spin structure of the nucleon
is briefly reviewed.
The main focus is on parton helicity distributions, orbital angular momentum
of partons as defined through generalized parton distributions, as well as 
single spin asymmetries and time-reversal odd correlation functions.}

\section{Introduction}
The history of the non-trivial nucleon spin structure started already 
in 1933 with the discovery of the anomalous magnetic moment of the 
proton by Frisch and Stern~\cite{frisch_33}.
This observation led to the important conclusion that the nucleon 
cannot be pointlike.

In the meantime the field has grown tremendously.
This short review concentrates on the QCD spin structure of the nucleon 
which is usually quantified in terms of various parton distributions.
In this context one is dealing with three kinds of parton distributions:
(1) forward distributions (quark and gluon helicity distribution), 
(2) generalized parton distributions (GPDs) which contain information 
on the orbital angular momentum of partons,
(3) transverse momentum dependent distributions (TMDs) which can lead
to single spin asymmetries (SSAs).
Related experiments are currently running at CERN, DESY, Jefferson Lab, 
and RHIC.

Many issues like the transversity distribution~\cite{barone_02}, 
parton distributions for $x \to 0,1$, various sum rules, subleading twist 
etc. cannot be covered. 
For such topics the reader is referred to existing review articles
(like, e.g., Refs.~\refcite{anselmino_95}--\refcite{bass_04}) and referenes
therein, as well as these proceedings.

\newpage
\section{Parton helicity distributions}

\subsection{Quark helicity distribution}
Up to now our knowledge about the quark helicity distribution $\Delta q$
has mostly come from inclusive lepton scattering off the nucleon.
By measuring double spin asymmetries (polarized lepton beam and polarized 
target) one can extract the structure function $g_1(x,Q^2)$ which is 
given by
\begin{equation}
g_1^{p,n} = \frac{1}{9} \Delta \Sigma
           \pm \frac{1}{12} \Delta q_3
           +   \frac{1}{36} \Delta q_8 \,,
\end{equation}
with the flavor combinations
\begin{eqnarray}
\Delta \Sigma & = & (\Delta u + \Delta \bar{u}) 
                  + (\Delta d + \Delta \bar{d}) 
                  + (\Delta s + \Delta \bar{s}) \,,
\nonumber \\
\Delta q_3 & = & (\Delta u + \Delta \bar{u}) 
               - (\Delta d + \Delta \bar{d}) \,,
\\
\Delta q_8 & = & (\Delta u + \Delta \bar{u}) 
               + (\Delta d + \Delta \bar{d}) 
             - 2 (\Delta s + \Delta \bar{s}) \,. 
\nonumber
\end{eqnarray}
In the past many QCD analyses, using (slightly) different assumptions 
and different schemes, of polarised DIS data were performed.
Information on the first moment of $\Delta q_3$ and $\Delta q_8$ from 
beta decay of the neutron and hyperons usually serves as an important 
independent constraint.
The results of such QCD analyses can roughly be summarized as follows:
while $\Delta \Sigma$ and $\Delta q_3$ are fairly well known, 
$\Delta q_8$ is not known with the same accuracy. 
In particular, this means that there still exists a considerable
uncertainty for the distribution of strange quarks.
Most importantly, however, inclusive DIS measurements do not permit
to determine $\Delta q$ and $\Delta \bar{q}$ separately.

\begin{figure}[ht]
\centerline{\epsfxsize=4cm\epsfbox{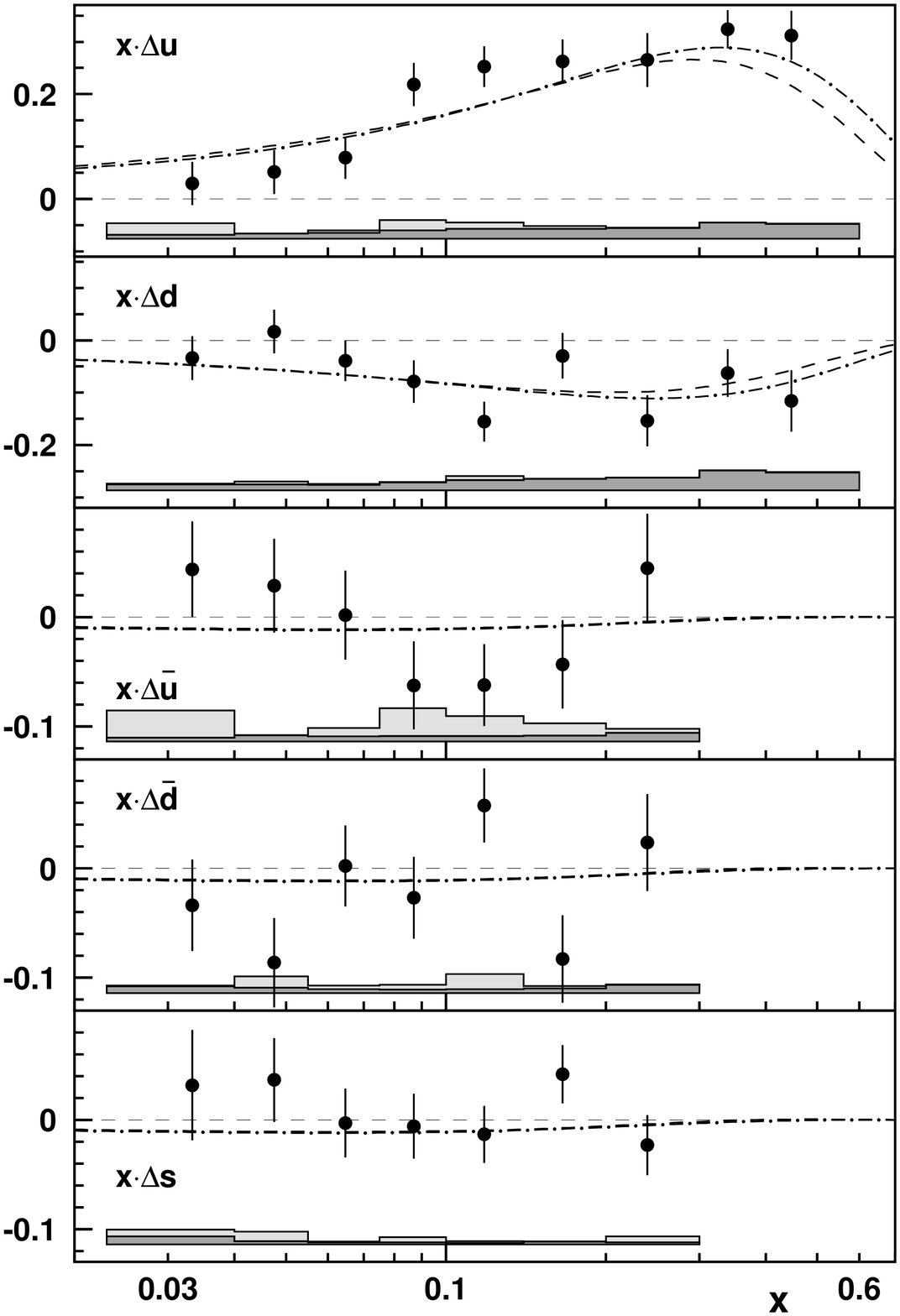} $\;\;$
\epsfxsize=5cm\epsfbox{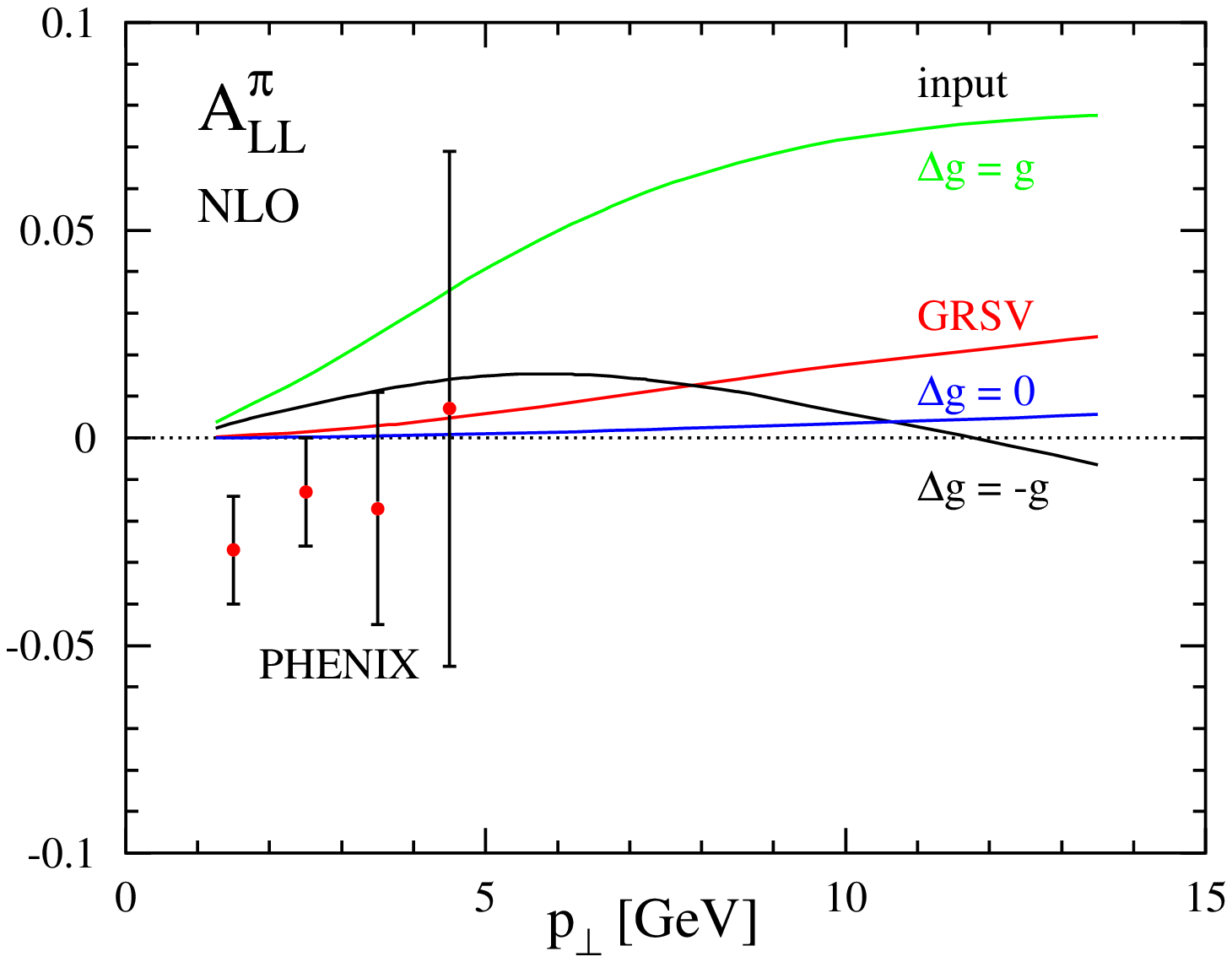}}   
\caption{Quark helicity distributions at 
 $\langle Q^2 \rangle = 2.5 \, \textrm{GeV}^2$
 from HERMES~{\protect \cite{HERMES_03}} (left panel),
 and longitudinal double spin asymmetry for $\vec{p} \vec{p} \to \pi^0 X$
 from PHENIX~{\protect \cite{PHENIX_04}} (right panel).  
\label{f:hel}}
\end{figure}

At this point additional information can be obtained from semi-inclusive 
DIS where one extracts the double spin 
asymmetry
\begin{equation} \label{e:as_sidis}
A^h \propto \frac{\sum_q \, e_q^2 \, \Delta q(x) \, D_q^h(z)}
           {\sum_q \, e_q^2 \, q(x) \, D_q^h(z)} \,.
\end{equation}
Detecting one hadron $h$ in the final state not only addresses the 
distribution of specific quark flavors (e.g., by looking at kaons  
one can learn something about the strange-quark distribution), but also 
makes it possible to separate the quark and antiquark distributions, 
since the fragmentation functions $D_q^h$, $D_{\bar{q}}^h$ put different 
weights on $\Delta q$ and $\Delta \bar{q}$.
The results for such an analysis from the HERMES 
Collaboration~\cite{HERMES_03,rubin_04} are shown in Fig.\ref{f:hel}, 
where, in particular, it turned out that the data are consistent
with a vanishing sea quark distribution for all three flavors.
It has been claimed, however, that the extraction method used in 
Ref.~\refcite{HERMES_03} has some model dependence~\cite{kotzinian_04}.
Recently, there has been quite some activity aiming at an entirely
model-independent analysis of semi-inclusive DIS 
data~\cite{christova_01,sissakian_04}.

Measuring a parity-violating SSA in $\vec{p} p \to W^{\pm} X$ 
can provide complementary information on the helicity distribution 
of the light flavors~\cite{craigie_83,bunce_00}, where for $W^+$ 
production one has
\begin{equation}
A_{L}^{W^+} = \frac{\Delta u(x_1) \bar{d}(x_2) - \Delta \bar{d}(x_1) u(x_2)}
                   {u (x_1) \bar{d}(x_2) + \bar{d}(x_1)u(x_2)} \,.
\end{equation}
Since $x_1$ and $x_2$ are fixed by external kinematics one can 
disentangle the contributions of the different flavors.
If, e.g., $x_1$ is large, then both $\Delta \bar{d}$ and $\bar{d}$
are small and one can extract the ratio $\Delta u / u$. 
On the other hand the ratio $\Delta \bar{d} / \bar{d}$ can be obtained 
if $x_2$ is large.
Analogously, production of $W^-$ allows one to measure $\Delta d / d$ 
and $\Delta \bar{u} / \bar{u}$.
This method is rather clean and is planned to be exploited at RICH. 
Eventually, polarized neutrino DIS could be used to get additional 
information on $\Delta s$ and $\Delta \bar{s}$~\cite{forte_04}.

\subsection{Gluon helicity distribution}
In order to get information on the gluon helicity distribution $\Delta g$ 
of the nucleon one studies lepton nucleon scattering as well as 
$pp$-collisions, where different final states are considered in both 
cases.

In the DIS measurements one tries to isolate the partonic subprocess of
photon-gluon fusion (PGF), $\gamma g \to q \bar{q}$.
From the experimental point of view, inclusive DIS represents the simplest 
reaction containing PGF.
However, since it only enters through evolution, this process merely 
provides a rather indirect measurement of $\Delta g$.
Because of the limited range in $x$ and $Q^2$, the currently available 
data put no strong constraint on $\Delta g$, even though a positive 
$\Delta g$ is prefered in the analyses. 
This is in contrast to the situation of the unpolarized gluon distribution, 
where a lot of information is coming from unpolarized inclusive DIS 
which has been explored in a wide kinematical region.

A more direct measurement of the PGF process is possible by detecting
high-$p_T$ jets or hadron pairs in the final state (see, e.g., 
Refs.~\refcite{anselmino_95}--\refcite{bass_04} and referenes therein).
In this context a special role is played by the production of a pair
of charmed mesons created through $\gamma g \to c \bar{c}$, because 
background processes like the QCD-Compton reaction $\gamma q \to g q$ 
are automatically suppressed without making specific kinematical cuts.
To measure $\Delta g$ via charm production is a central aim of the
COMPASS Collaboration~\cite{COMPASS_96,schill_04}.

The only published numbers for $\Delta g$ from such type of reactions
are coming from the production of high-$p_T$ hadron pairs.
The measurements of the HERMES~\cite{HERMES_00} and SMC~\cite{SMC_04} 
Collaborations, performed at different average values of $x$, yielded 
\begin{eqnarray}
\Delta g / g |_{\langle x \rangle = 0.17} & = & 
 0.41 \pm 0.18 \, \textrm{(stat)} \pm 0.03 \, \textrm{(syst)}\qquad
 \textrm{(from Ref.~\refcite{HERMES_00})} \,,
\\
\Delta g / g |_{\langle x \rangle = 0.7} & = & 
 -0.20 \pm 0.28 \, \textrm{(stat)} \pm 0.10 \, \textrm{(syst)}\qquad
 \textrm{(from Ref.~\refcite{SMC_04})} \,.
\end{eqnarray}
Unfortunately, these data are still suffering from large statistical 
errors. 
While the SMC result was obtained in the DIS regime 
$(Q^2 > 1\, \textrm{GeV}^2)$, HERMES used photoproduction which led to 
speculations about background contributions from resolved photons.

The second class of processes providing information on $\Delta g$ are 
longitudinal double spin asymmetries in proton-proton collisions.
To be specific, the following reactions are considered: 
prompt photon production ($\vec{p}\vec{p} \to \gamma X$), 
production of heavy flavors ($\vec{p}\vec{p} \to c \bar{c} X, \, b \bar{b} X$), 
jet production ($\vec{p}\vec{p} \to \textrm{jet} X$), 
as well as inclusive production of hadrons ($\vec{p}\vec{p} \to h X$). 
The processes have already been computed up to NLO in QCD.
A detailed discussion of the advantages and drawbacks of the different
reactions can be found in 
Refs.~\refcite{bunce_00,fillipone_01,jaeger_03,jaeger_04a,jaeger_04b} 
and references therein.
At RICH there are extensive ongoing activities in order to study
the various channels for different kinematics.

The first published data are from the PHENIX Collaboration for inclusive 
production of neutral pions~\cite{PHENIX_04,fukao_04}.
The asymmetry is shown in Fig.\ref{f:hel} as function of the transverse
momentum of the pion, and compared to a NLO calculation~\cite{jaeger_03}.
Measuring $A_{LL}^{\pi}$ with good statistics at higher values of 
$p_{\perp}$, where the sensitivity of the asymmetry to the gluon helicity
is larger as compared to the low $p_{\perp}$ region, can already provide 
an important constraint on $\Delta g$.

\section{Generalized parton distributions and orbital angular momentum}
Knowing the helicity distributions is not sufficient to understand how 
the spin of the nucleon is decomposed.
One also needs information on the orbital angular momentum of the partons.
In 1996 it was shown~\cite{ji_97a} that generalized parton distributions
(see, e.g., Refs.~\refcite{mueller_94}--\refcite{garcon_04}) can provide 
the pertinent information.
GPDs appear in the description of hard exclusive processes like 
deep-virtual Compton scattering off the nucleon and meson production,
where in both cases data have already been published 
(see Refs.~\refcite{diehl_03,HERMES_04b,CLAS_04b} and references therein).
Neglecting the scale dependence, GPDs are functions of three variables,
$x$, $\xi$, $t$.
While $\xi$ and $t$, describing the longitudinal and total momentum 
transfer to the nucleon, are fixed by the external kinematics of an 
experiment, $x$ is integrated over which complicates the extraction
of the $x$-dependence of GPDs.

GPDs contain a vast amount of physics, and show several interesting 
properties which put strong constraints on models.
They are related to forward parton distributions and nucleon form factors, 
obey the so-called polynomiality condition~\cite{ji_98}, and satisfy 
positivity bounds~\cite{pobylitsa_02}. 
Moreover, they contain information on the shear forces partons experience 
in the nucleon~\cite{polyakov_03}.
In particular, they can provide a 3-dimensional picture of the 
nucleon~\cite{burkardt_03}.

Concerning the nucleon spin structure it is important that the total 
angular momentum (for longitudinal polarization) of quarks is related to 
the GPDs according to~\cite{ji_97a}
\begin{equation}
J_{q}^{z} = \frac{1}{2} \int_{-1}^{1} dx x
 \Big[ H_{q}(x,\xi,t=0) + E_{q}(x,\xi,t=0) \Big] \,,
\end{equation}
where $H_{q}(x,0,0) = q(x)$, while the GPD $E_{q}$ has no relation to a 
normal forward distribution.
For $J_{g}^{z}$ an analogous formula holds.
Knowing both the total angular momentum and the helicity of partons allows 
one to address the orbital angular momentum by means of the decomposition
 \begin{equation} \label{e:decomp}
\frac{1}{2} = \sum_q J_{q}^{z} + J_{g}^{z}
 =  \sum_q \bigg[ \frac{1}{2} \int_{0}^{1} dx 
          \Big( \Delta q(x) + \Delta \bar{q} (x) \Big) + L_{q}^{z} \bigg]
    + \Delta g + L_{g}^{z} \,.
\end{equation}
(Note that also for a transversely polarized nucleon a decomposition like 
in~(\ref{e:decomp}) has been proposed~\cite{bakker_04}.)
Recently, Lattice QCD~\cite{LHP_03,QCDSF_04} as well as models and 
phenomenlogical parametrizations of GPDs~\cite{diehl_04,guidal_04,ossmann_04} 
were used to estimate the orbital angular momentum of the quarks.
Lattice data, e.g., result in a small contribution to the angular momentum
if one sums over the quarks, but the uncertainties of these calculations 
are still large.

\section{Single spin asymmetries}
\begin{figure}[ht]
\centerline{\epsfxsize=5cm\epsfbox{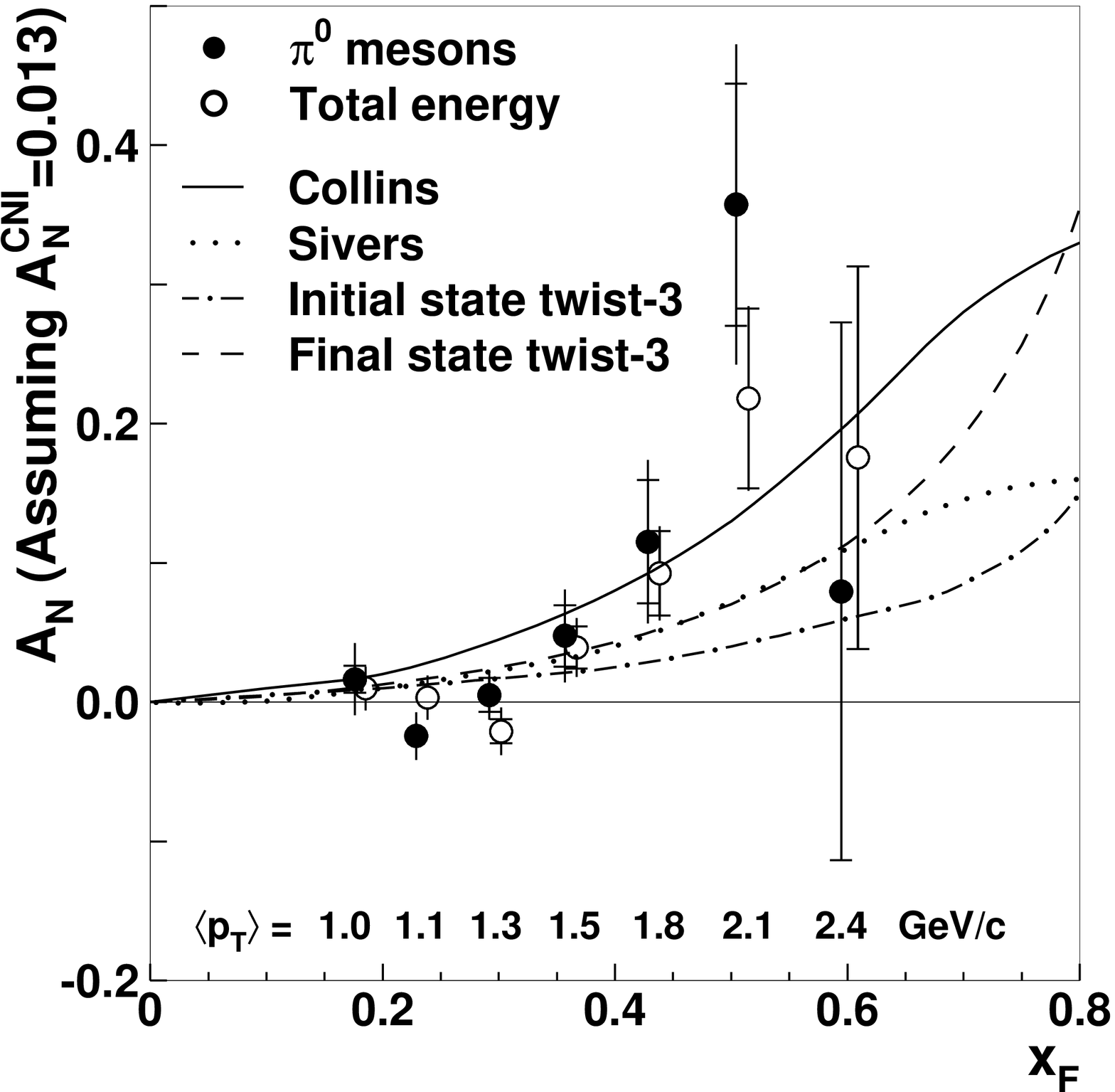} \\
\epsfxsize=5cm\epsfbox{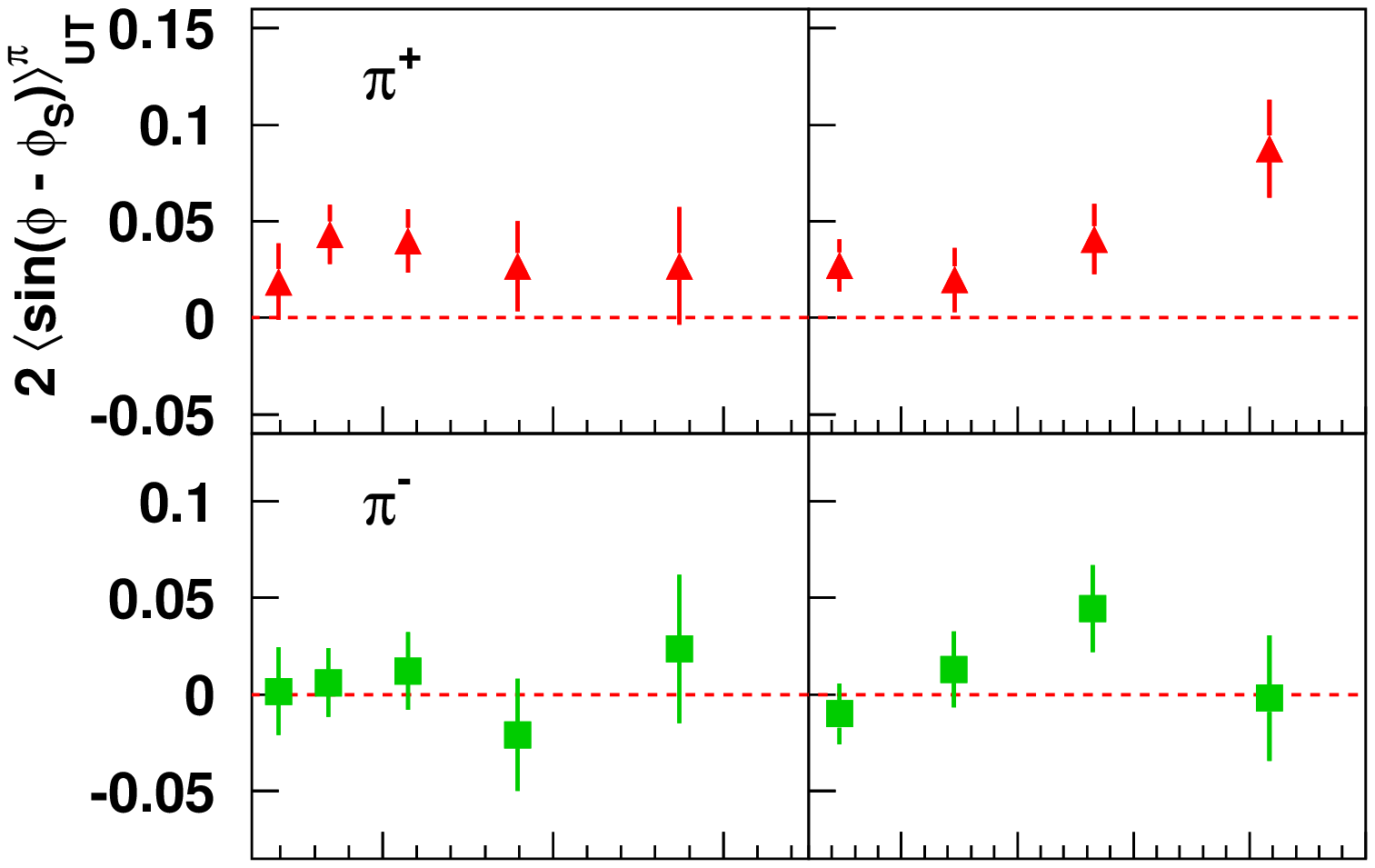}}   
\caption{Transverse single spin asymmetries: $A_N$ in 
 $p^{\uparrow} p \to \pi^0 X$ from STAR~{\protect \cite{STAR_04}} (left panel), 
 and Sivers asymmetry in $e p^{\uparrow} \to e \pi X$ as function of 
 $x$ ({\it lhs}) and $z$ ({\it rhs}) from 
 HERMES~{\protect \cite{HERMES_04}}(right panel).  
\label{f:ssa}}
\end{figure}

Single spin asymmetries are currently under intense investigation from 
both the experimental and theoretical point of view.
For the process $p^{\uparrow} p \to \pi X$, e.g., Fermi-Lab~\cite{FLAB_91} 
observed large transverse SSAs (up to 40\%) at the {\it cm}-energy
$\sqrt{s} = 20 \, \textrm{GeV}$, and recent results from the STAR 
Collaboration~\cite{STAR_04} have shown that the effect survives 
at $\sqrt{s} = 200 \, \textrm{GeV}$ (see Fig.\ref{f:ssa}).
Also for pion production in semi-inclusive DIS non-vanishing transverse
SSAs have been observed~\cite{HERMES_04} (see Fig.\ref{f:ssa}).

In general, SSAs are generated by so-called time-reversal odd (T-odd)
correlation functions (parton distributions and fragmentation functions).
They vanish in leading twist collinear factorization~\cite{kane_78}.
To get non-zero effects one has to resort to (collinear) 
twist-3 correlators~\cite{efremov_85,qiu_91} or to transverse momentum 
dependent functions~\cite{ralston_79,mulders_96,boer_98}.
There exist four T-odd leading twist TMD correlation functions, where the 
Sivers function $f_{1T}^{\perp}$~\cite{sivers_90}, describing the azimuthal 
asymmetry of quarks in a transversely polarized target, is the most 
prominent T-odd parton distribution.
In the case of fragmentation the Collins function~\cite{collins_93} 
(transition of a transversely polarized quark into an unpolarized hadron) 
has attracted a lot of interest, since in semi-inclusive DIS it gets coupled 
to the transversity distribution of the nucleon.

For $A_N$ in $pp$-collisions both TMD twist-2 and collinear twist-3 correlators 
were used to describe the data as can be seen in Fig.~\ref{f:ssa}.
For the twist-2 analysis, very recently the invoked kinematics has been revisited
carefully.
As a result it turns out that the Collins mechanism actually cannot explain 
the data~\cite{anselmino_04b}, while the Sivers mechanism could well do 
so~\cite{dalesio_04}.
In contrast to $A_N$, in semi-inclusive DIS at low transverse momentum of 
the detected hadron one can unambiguously select the Sivers mechanism shown
in Fig.~\ref{f:ssa}.

For quite some time it was believed that T-odd TMD distributions like the 
Sivers function should vanish because of T-invariance of the strong 
interaction~\cite{collins_93}, whereas T-odd fragmentation functions may well 
exist because of final state interactions~\cite{collins_93,bacchetta_01}.
However, in 2002 a simple spectator model calculation provided a non-zero
SSA in DIS~\cite{brodsky_02}.
A reanalysis then revealed that in fact the Sivers function can be non-zero, 
but only if the Wilson-line ensuring color gauge invariance is taken into
account in the operator defintion~\cite{collins_02}.
The presence of the Wilson line which can be process-dependent in turn 
endangers universality of TMD correlation 
functions~\cite{collins_02,boer_03,bomhof_04}.
This problem affects also the soft factor appearing in factorization 
formulae for transverse momentum dependent processes.
The schematical structure of the factorization formula for semi-inclusive 
DIS is~\cite{collins_81,ji_04,collins_04}
\begin{equation}
 \sigma_{DIS} \propto \textrm{pdf} \times \textrm{frag} \times \hat{\sigma}_{part} 
 \times \textrm{soft} \,.
\end{equation}
For unintegrated Drell-Yan and $e^+ e^- \to h_1 h_2 X$ if the two hadrons 
are almost back-to-back one is dealing with corresponding formulae.
While time-reversal can be used to relate parton distributions in DIS which 
contain future-pointing Wilson lines to distributions in Drell-Yan with 
past-pointing lines, this is not possible for fragmentation functions.
Nevertheless, by considering the analytic properties of the fragmentation
correlator, it can be shown that fragmentation functions are 
universal~\cite{collins_04,metz_02}.
This result, in particular, justifies to relate the Collins function in
$e^+ e^-$-annihilation and semi-inclusive DIS~\cite{efremov_01,seidl_04}.
Also for the soft factor universality between the three mentioned processes 
can be established~\cite{collins_04}.
Only T-odd parton distributions are non-universal in the sense that they
have a reversed sign in DIS as compared to Drell-Yan, i.e.,
\begin{equation}
f_{1T}^{\perp} \Big |_{DY} = -  f_{1T}^{\perp} \Big |_{DIS} \,.
\end{equation}
This relation should be checked experimentally.

There are many more interesting developments in the field of SSAs.
For instance, a relation between the sign of the Sivers function and
the anomalous magnetic moment of a given quark flavor was 
given~\cite{burkardt_02}.
Moreover, a sum rule relating the Sivers effect for quarks and gluons
was derived~\cite{burkardt_04}.
It was also proposed to measure the gluon Sivers function through
jet correlations in $p^{\uparrow} p$-collisions~\cite{boer_04}, 
and charm production ($p^{\uparrow} p \to D X$)~\cite{anselmino_04a}.

\section{Conclusions}
We have briefly reviewed the status of the QCD spin structure of the 
nucleon.
Currently, an enormous amount of activities is dealing with this vast 
and very interesting field.

Historically, the first subject which was studied intensely is the
physics of parton helicity distributions, and today we already have 
a considerable knowledge about the quark helicity distribution.
Uncertainties still exist in the strange quark sector and in the 
separation of valence and sea quark distributions, but many current 
activities are aiming at an improvement of this situation.
In contrast to $\Delta q$, the gluon helicity distribution is still just
weakly constrained.
Nevertheless, a lot of new information, which is supposed to come in the 
near future from COMPASS and the various measurements at RHIC, will 
certainly increase our knowlegde about $\Delta g$.

Also generalized parton distributions can provide important information
in order to resolve the spin puzzle of the nucleon, because the orbital
angular momentum of partons is related to these objects.
Using Lattice QCD as well as phenomenological approaches people have 
exploited this connection to determine the orbital angular momentum 
of quarks.
At present, the situation is not yet conclusive, but should definitely 
improve in the future.
In particular, many new preliminay data for hard exclusive reactions 
on the nucleon from COMPASS, HERMES, and Jefferson Lab exist.
These data will also help to clarify the role played by orbital 
angular momentum in the spin sum rule of the nucleon.

The discovery that time-reversal odd parton distributions in general are
non-zero gave a strong boost to the interesting subject of single spin 
asymmetries over the past three years.
Since then a lot of progress has been made on both the theoretical but also 
the experimental side.
In this context it has been a crucial discovery that the presence of the 
Wilson line in transverse momentum dependent correlation functions 
is mandatory.
Because this field in some sense is still rather young, more fundamental 
results are to be expected.
The large amount of already existing, preliminary, and forthcoming data 
from lepton-nucleon and proton-proton collisions will further improve our 
understanding of the origin of single spin asymmetries.


\end{document}